\documentclass[prc,aps,twocolumn,nofootinbib,showpacs,showkeys]{revtex4}
\usepackage{graphicx,epsf,amsfonts,amssymb,amsbsy}
\begin{document}

\title{Hybrid stars with color superconductivity within a nonlocal chiral
quark model}
\author{H. ~Grigorian}
\email{hovik@darss.mpg.uni-rostock.de}
\affiliation{Fachbereich Physik, Universit\"at Rostock, D-18051
Rostock, Germany\\
Department of Physics, Yerevan State University,
375025 Yerevan, Armenia
}
\author{D. ~Blaschke}
\email{david.blaschke@physik.uni-rostock.de}
\affiliation{Fachbereich Physik, Universit\"at Rostock, D-18051
Rostock, Germany\\
Bogoliubov  Laboratory of Theoretical Physics,
Joint Institute for Nuclear Research, 141980 Dubna, Moscow Region,
Russia}
\author{D. ~N. ~Aguilera}
\email{deborah@darss.mpg.uni-rostock.de}
\affiliation{Fachbereich Physik, Universit\"at Rostock, D-18051
Rostock, Germany\\
Instituto de F\'{\i}sica Rosario, Bv. 27 de febrero 210 bis,
2000 Rosario, Argentina}

\begin{abstract}
The equation of state for quark matter is derived for a nonlocal,
chiral quark model within the mean field approximation. Special
emphasis is on the occurrence of a diquark condensate which
signals a phase transition to color superconductivity and its
effects on the equation of state. We present a fit formula for the
Bag pressure, which is densitiy dependent in case when the quark matter
is color superconducting. We calculate the quark star
configurations by solving the Tolman- Oppenheimer- Volkoff
equations and demonstrate the effects of diquark condensation on
the stability of hybrid stars for different formfactors of the
quark interaction.
\end{abstract}

\maketitle

\section{Introduction}
The investigation of color superconductivity in quark matter
\cite{Barrois:1977xd,Bailin:bm} which was revived by applying nonperturbative
QCD motivated interactions \cite{Rapp:1997zu,alford99} finds most of its
justification in the possible importance for the
physics of compact star interiors \cite{Blaschke:uj} and related
observable phenomena like neutron star cooling
\cite{Blaschke:1999qx,Page:2000wt,Blaschke:2000dy},
gamma-ray bursts
\cite{Hong:2001gt,Ouyed:2001bm,Aguilera:2002dh,Blaschke:2003yn},
gravitational wave signals for compact stars mergers \cite{Oechslin:2004yj}
and others \cite{Rajagopal:jq}.
Since calculations of quark pairing predict
values of the energy gap $\Delta \sim 100$ MeV and corresponding critical
temperatures for the phase transition to the superconducting state are
expected to follow the BCS relation $T_c = 0.57 ~\Delta$ for spherical
wave pairing,
quark matter in compact stars should always be in one of the superconducting
phases.

The question arises whether conditions in compact stars allow the occurence
of quark matter and the formation of stable configurations of hybrid stars.
In order to give an answer to this question one has to rely on models for
the equation of state which necessarily introduce free parameters and therefore
some arbitrariness in the results \cite{Alford:2002rj}.
In particular, the question whether color superconductivity shall be realized
in the 2-Flavor Superconductivity (2SC) or Color Flavor Locking (CFL) phase in the compact star interior has been discussed
controversely \cite{Alford:2002kj,Steiner:2002gx}.
We will restrict our further discussion to dynamical models of the NJL type
with their parameters adjusted by fitting hadron properties in the vacuum
before extrapolating to finite temperatures and densities within the Matsubara
formalism.

Within those models it has been shown that strange quark matter
occurs only at densities well above the deconfinement transition, for
chemical potentials which are barely reached in the very center of a compact
star
\cite{Gocke:2001ri,Buballa:2001gj,Neumann:2002jm,Shovkovy:2003ce,Baldo:2002ju}.
The interesting and much investigated CFL phase could thus play only
a marginal role for the physics of compact stars.
The stability of the so obtained hybrid star configurations, however, appears
to depend sensitively on details of the model, including the hadronic
phase.

In the present paper we are investigating this dependence in a
systematic way by employing a nonlocal chiral quark model which
allows to vary the formfactor of the interaction kernel while
describing the same set of hadronic vacuum properties. We provide
a polynomial fit formula of our quark matter equation of state
(EoS) which proves useful for applications to compact star
phenomenology, as e.g. the cooling \cite{Grigorian:2003} and
rotational evolution \cite{Poghosyan:2000mr} or the merging
\cite{Oechslin:2004yj} of neutron stars. In order to compare the
results of the present work for hybrid star configurations with
observational constraints, we pick the example of the compact
object RX J$185635-3754$, for which limits for both the mass and
the radius have been reported \cite{Prakash:2002xx,Zane:2003bp}. A
further restriction in the mass - radius plane of possible stable
configurations comes from the constraint given by the surface
redshift measurement of EXO 0748-676 \cite{Cottam:2002cu}.

\section{Equation of state of hybrid star matter in $\beta$ equilibrium}

\subsection{Quark matter with color superconductivity}

We consider the grand canonical thermodynamic potential
for 2SC quark matter within a nonlocal chiral quark model
\cite{Blaschke:2003yn}
where in the mean field approximation the mass gap $\phi_f$ and the
diquark gap $\Delta$ appear as order parameters and a decomposition
into color ($c\in\{r,b,g\}$) and flavor ($f\in\{u,d\}$) degrees of
freedom can be made.

\begin{eqnarray}
&&
\Omega_q(\{\phi_f\},\Delta;\{\mu_{fc}\},T)=
\sum_{c,f}\Omega^{c,f}(\phi_f,\Delta;\mu_{fc},T)~,
\end{eqnarray}
where  $T$ is the  temperature and $\mu_{fc}$
the chemical potential for the quark with flavor $f$ and color $c$.

The contribution of quarks with  given color $c$ and flavor $f$ to the
thermodynamic potential is
\begin{eqnarray}
&&
\Omega^{c,f}(\phi_f,\Delta;\mu_{fc},T)+ \Omega_{vac}^c=
\frac{\phi_f^2}{24~G_1}+\frac{\Delta^2}{24~G_2}
\nonumber\\
&&-
\frac{1}{\pi^2}
\int^\infty_0dqq^2
\{
\omega\left[
\epsilon_c(E_f(q)+\mu_{fc}),T
\right]
+
\nonumber\\
&&
\omega\left[
\epsilon_c(E_f(q)-\mu_{fc}),T
\right]
\}~,
\label{ome2}
\end{eqnarray}
where $G_{1}$ and $G_{2}$ are coupling constants in the
scalar meson and diquark channels, respectively.
The dispersion relation for unpaired quarks with dynamical mass
function $m_f(q)=m_f+g(q)\phi_f$
is given by
\begin{eqnarray}
&& E_f(q)=
\sqrt{q^2+m^2_f(q)}~.
\end{eqnarray}
In Eq. (\ref{ome2}) we have introduced the notation
\begin{eqnarray}
&&
\omega\left[\epsilon_c,T\right]= T\ln\left
[1+\exp\left(-\frac{\epsilon_c}{T}\right)\right]+\frac{\epsilon_c}{2}~,
\label{ome3}
\end{eqnarray}
where the first argument is  given by
\begin{eqnarray}
&&\epsilon_c(\xi)=\xi\sqrt{1+\Delta^2_c/\xi^2}~.
\end{eqnarray}
When we choose the green and blue colors to be paired and the red
ones to remain unpaired, we have
\begin{eqnarray}
&& \Delta_c=g(q)\Delta(\delta_{c,b}+\delta_{c,g}).
\end{eqnarray}

For a homogeneous system in equilibrium, the minimum of the
thermodynamic potential $\Omega_q$ with respect to the order
parameters $\{\phi_f\}$ and $\Delta$ corresponds to a negative
pressure; therefore the constant
$\Omega_{vac}=\sum_{c}\Omega_{vac}^c$ is chosen such that the
pressure of the physical vacuum vanishes.

The nonlocality of the interaction between the quarks
in both channels $q\bar{q}$ and $qq$
is implemented via the same formfactor functions
$g(q)$ in the momentum space.
In our calculations we use the Gaussian (G), Lorentzian (L) and cutoff (NJL)
type formfactors defined as
\begin{eqnarray}
\label{GF}
g_{\rm G}(q) &=& \exp(-q^2/\Lambda_{\rm G}^2)~,\\
\label{LF}
g_{\rm L}(q) &=& [1 + (q/\Lambda_{\rm L})^2]^{-1},\\
\label{NF}
g_{\rm NJL}(q) &=& \theta(1 - q/\Lambda_{\rm NJL})~.
\end{eqnarray}
The parameter sets (quark mass $m$, coupling constant $G_1$, interaction range
$\Lambda$) for the above formfactor models (see Tab. \ref{par}) are
fixed by the pion mass $m_{\pi}=140$  MeV, pion decay constant $f_{\pi}=93$
MeV and the constituent quark mass $m_0=330$ MeV
at $T=\mu=0$  \cite{Schmidt:di}.
The diquark coupling constant $G_2$ is a free parameter of the approach which we vary as
$G_2=\eta~ G_1$.

Following Ref. \cite{Huang:2002zd} we introduce
the quark chemical potential for the color $c$, $\mu_{qc}$
and the chemical potential of the isospin asymmetry, $\mu_I$, defined as

\begin{eqnarray}
\mu_{qc}&=&(\mu_{uc}+\mu_{dc})/2\\
\mu_I&=&(\mu_{uc}-\mu_{dc})/2,
\label{muqmuI}
\end{eqnarray}

where the latter is color independent.

The diquark condensation in the 2SC phase induces a color asymmetry
which is proportional to the chemical potential $\mu_8$. 
Therefore we can write
\begin{eqnarray}
\mu_{qc}=\mu_q+\frac{\mu_8}{3}(\delta_{c,b}+\delta_{c,g}-2\delta_{c,r})~,
\label{mu8}
\end{eqnarray}
where $\mu_q$ and $\mu_8$ are conjugate to the quark number
density and the color charge density, respectively.

As has been shown in \cite{Frank:2003ve} for the 2SC phase the relation
$\phi_u=\phi_d=\phi$ holds so that the quark thermodynamic potential is
\cite{Kiriyama:2001ud}

\begin{eqnarray} \label{Omeg1}
&&\Omega_q(\phi,\Delta;\mu_q,\mu_I,\mu_8,T)+\Omega_{vac}
=\frac{\phi^2}{4G_1}+\frac{\Delta^2}{4G_2}
\nonumber\\
& &
-\frac{1}{\pi^2}\int^\infty_0dqq^2\{
\omega\left[\epsilon_r(-\mu_q+\frac{2}{3}\mu_8-\mu_I),T\right]+
\nonumber\\
& &
\omega\left[\epsilon_r(\mu_q-\frac{2}{3}\mu_8-\mu_I),T\right]+
\omega\left[\epsilon_r(-\mu_q+\frac{2}{3}\mu_8+\mu_I),T\right]+
\nonumber\\
& &
\omega\left[\epsilon_r(\mu_q-\frac{2}{3}\mu_8+\mu_I),T\right]
\}\nonumber\\
& &
-\frac{2}{\pi^2}\int^\infty_0dqq^2\{
\omega\left[\epsilon_b(E(q)-\mu_q-\frac{1}{3}\mu_8)-\mu_I,T\right]+
\nonumber\\
& &
\omega\left[\epsilon_b(E(q)+\mu_q+\frac{1}{3}\mu_8)-\mu_I,T\right]+
\nonumber\\
& &
\omega\left[\epsilon_b(E(q)-\mu_q-\frac{1}{3}\mu_8)+\mu_I,T\right]+
\nonumber\\
& &
\omega\left[\epsilon_b(E(q)+\mu_q+\frac{1}{3}\mu_8)+\mu_I,T\right]
\}~,
\label{ome9}
\end{eqnarray}

where the factor $2$ in the last integral comes from the degeneracy of the
blue and green colors ($\epsilon_b=\epsilon_g$).

The total thermodynamic potential $\Omega$ contains besides the quark
contribution
$\Omega_q$ also that of the leptons $\Omega^{id}$
\begin{eqnarray}
\label{omega_tot}
\Omega(\phi,\Delta;\mu_q,\mu_I,\mu_8,\mu_l,T)&=&
\Omega_q(\phi,\Delta;\mu_q,\mu_I,\mu_8,T) \nonumber  \\
&+& \sum_{l \in \{e,\bar \nu_e, \nu_e \}} \Omega^{id}(\mu_l,T)~.
\end{eqnarray}

where latter are assumed to be a massless, ideal Fermi gas
\begin{eqnarray}
&&
\Omega^{id}(\mu,T)=-\frac{1}{12\pi^2}\mu^4-\frac{1}{6}\mu^2T^2-\frac{7}{180}
\pi^2T^4~.
\end{eqnarray}

At the present stage, we do include only contributions of the
first family of leptons in the thermodynamic potential.

The conditions for the local extremum of $\Omega_q$ correspond to
coupled gap equations for the two order parameters $\phi$ and
$\Delta$
\begin{eqnarray}
&&
{\partial \Omega \over \partial \phi}\bigg|_{\phi=\phi_0,\Delta=\Delta_0}=
{\partial \Omega \over \partial \Delta}\bigg|_{\phi=\phi_0,\Delta=\Delta_0}=0~.
\label{GapEq}
\end{eqnarray}
The global minimum of $\Omega_q$ represents the state of
thermodynamic equilibrium from which all equations of state can be
obtained by derivation.

\subsection{Beta equilibrium, charge and color neutrality}

The stellar matter in equilibrium has to obey the constraints of
$\beta$-equilibrium ($d \longrightarrow u+e^-+\bar \nu_e$, $u + e^-
\longrightarrow d + \nu_e$), expressed as
\begin{eqnarray}
\mu_{dc}=\mu_{uc}+\mu_e~,
\label{betaEq}
\end{eqnarray}
color and electric charge neutrality and baryon number conservation.

We use in the following the electric charge density
\begin{equation}
\label{Q}
Q = \frac{2}{3}\sum_c n_{uc}-\frac{1}{3}\sum_c n_{dc}-n_e ~,
\end{equation}
the baryon number density
\begin{equation}
\label{nB}
n_B = \frac{1}{3}\sum_{f,c}n_{fc}~,
\end{equation}
and the color number density
\begin{equation}
n_8=\frac{1}{3}\sum_f(n_{fb}+n_{fg}-2n_{fr})~.
\label{n8}
\end{equation}
The number densities $n_j$ occuring on the right hand sides of the above
Eqs. (\ref{Q}) - (\ref{n8}) are
defined as derivatives of the thermodynamic potential (\ref{omega_tot})
with respect to corresponding chemical potentials $\mu_j$
\begin{eqnarray}
n_j=-\frac{\partial \Omega}{\partial \mu_j}\bigg|_{\phi_0,\Delta_0;T,\{\mu_j,
j\not=i\}}~,
\end{eqnarray}
Here the index $j$ stands for the particle species.

In order to express the Gibbs free enthalpy density $G$ in terms
of those chemical potentials which are conjugate to the conserved 
densities and to implement the $\beta-$equilibrium condition (\ref{betaEq}) 
we make the following algebraic transformations
\begin{eqnarray}
G &=& \sum_{f,c}\mu_{fc}n_{fc} + \mu_en_e \nonumber\\
&=&\frac{1}{3}\sum_c (3\mu_{qc}-\mu_I) (n_{dc}+n_{uc}) -\mu_eQ\nonumber\\
&=& \mu_B~n_B +\mu_Q~ Q + \mu_8~n_8~,
\label{Gibbs1}
\end{eqnarray}
where we have defined the chemical potential $\mu_B=3\mu_{q}-\mu_I$ 
conjugate to the baryon number density $n_B$ in the same way as  
$\mu_Q = -\mu_e$ is the chemical potential conjugate to
$Q$ and  $\mu_8$ to $n_8$.
Then, the electric and color charge neutrality conditions read,
\begin{eqnarray}
Q &=& 0~,\\
n_8 &=& 0~,
\end{eqnarray}
at given  $n_B$.
The solution of the gap equations (\ref{GapEq})
can be performed under these constraints.

The solution of the color neutrality condition shows that
$\mu_8$ is about $5\div 7$ MeV in the region of relevant densities
($\mu_q\simeq 300\div 500$ MeV). Since $\mu_I$ is independent
of $\mu_8$ we consider $\mu_{qc}\simeq\mu_q$ ($\mu_8\simeq0$) in our 
following calculations.

\begin{figure}[htb]
  \begin{center}
    \includegraphics[width=0.65\linewidth,angle=-90]{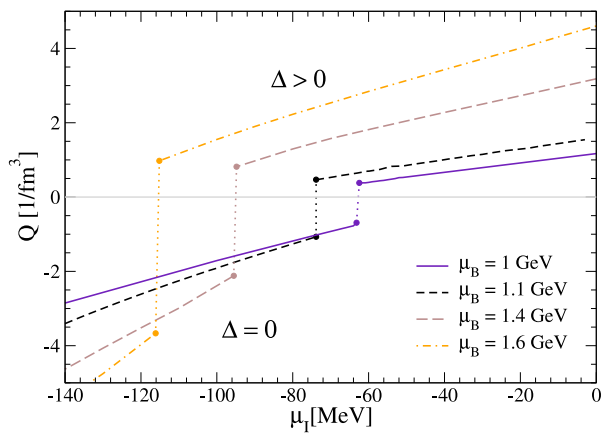}
    \vspace*{0.7cm}
    \caption{Electric charge density  for the 2SC ($\Delta>0$)
and normal ($\Delta=0$) quark matter phases
 as a function of $\mu_I$ for different fixed values of $\mu_B$.
The end points of the lines for given $\mu_B$ denote states with the 
same pressure and represent subphases in the Glendenning construction.
    \label{fig:Qanalisis}}
  \end{center}
\end{figure}

To demonstrate how to define a charge neutral state of  quark
matter in 2SC phase we plot  
in Fig. \ref{fig:Qanalisis} the electric charge density $Q$ 
as a function of $\mu_I$  for different fixed values 
of $\mu_B$, when system is in the global minimum of the thermodynamic
potential and $\eta =1$. 
As has been shown before in Ref. \cite{Neumann:2002jm} for the
NJL model case, the pure phases ($\Delta>0$: superconducting,  $\Delta=0$: 
normal) in general are charged. 
These branches end at critical values of $\mu_I$ where their pressure 
is equal and the corresponding states are degenerate
\begin{eqnarray}
P = P_{\Delta=0}(\mu_B,\mu_I,\mu_e,T)=P_{\Delta>0}(\mu_B,\mu_I,\mu_e,T)~.
\label{Pressure}
\end{eqnarray}
In order to fulfill the charge neutrality condition one can 
construct a homogeneous mixed phase of these states using the Gibbs 
conditions  \cite{Glendenning:1992vb}. 

\begin{figure}[htb]
  \begin{center}
    \includegraphics[width=0.65\linewidth,angle=-90]{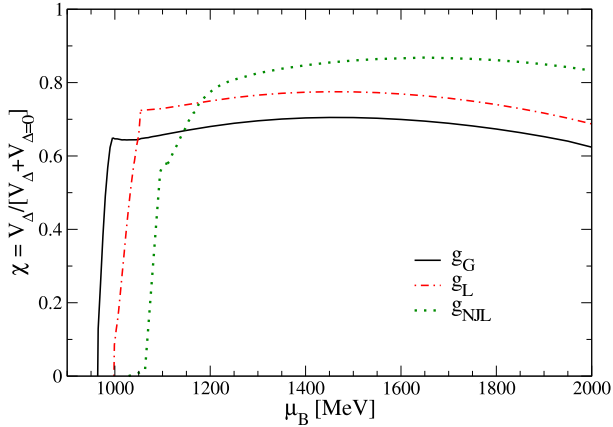}
    \vspace*{0.7cm}
    \caption{Volume fraction $\chi$ of the phase with nonvanishing
      diquark condensate obtained by a Glendenning construction of
a charge-neutral mixed phase. Results are shown for three different
formfactors introduced in the text.}
    \label{fig:volfraction}
  \end{center}
\end{figure}

The volume fraction that is occupied by the subphase with diquark condensation
is defined by the charges in the subphases
\begin{eqnarray}
\chi = Q_{\Delta>0}/(Q_{\Delta>0}-Q_{\Delta=0})
\label{Charge}
\end{eqnarray}
and is plotted in Fig. \ref{fig:volfraction} for the different formfactor
functions as a function of  $\mu_B$.

In the same way, the  number densities for the different particle species $j$
and the energy density are given by
\begin{eqnarray}
n_j = \chi n_{j_{\Delta>0}}+ (1-\chi)n_{j_{\Delta=0}}\\
\varepsilon = \chi \varepsilon_{\Delta>0}+ (1-\chi)\varepsilon_{\Delta=0}~.
\label{Number}
\end{eqnarray}

The formulas (\ref{Pressure})-(\ref{Number}) define a complete set of 
thermodynamic relations  and can be evaluated numerically.
In the next section we present the results in a form analog to the
Bag model which has been widely used in the phenomenology of quark matter.

\section{Quark star EoS and fit formulas}

\begin{figure}[htb]
  \begin{center}
    \includegraphics[width=0.65\linewidth,angle=-90]{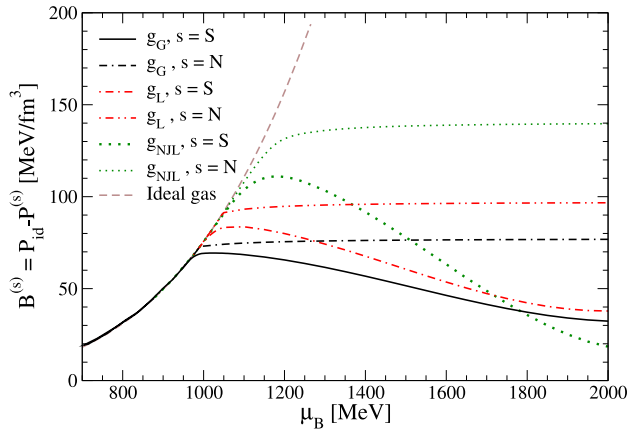}
    \vspace*{0.7cm}
    \caption{Bag pressure for different formfactors of the quark interaction
in dependence on the baryon chemical potential for  $\eta = 0$ and for 
 $\eta = 1$. For the latter the superconducting phase is realized.}
    \label{fig:bagnew}
  \end{center}
\end{figure}

We calculate the quark matter EoS within this nonlocal chiral model
\cite{Blaschke:2003yn} and display the results for the pressure in a
form remniscent of a bag model
\begin{equation}
\label{press}
  P^{(s)} = P_{id}(\mu_B) - B^{(s)}(\mu_B)~,
\end{equation}
where $P_{id}(\mu_B)$ is the ideal gas pressure of quarks and $B^{(s)}(\mu_B)$ a
{\it density dependent} bag pressure, see Fig. \ref{fig:bagnew}. 
The occurrence of diquark condensation depends on the value of $\eta=G_2/G_1$ and 
the superscript $s\in\{S,N\}$ indicates whether we
consider  the matter in the  superconducting mixed phase ($\eta=1$) or
in the normal phase ($\eta=0$), respectively.


According to heuristic expectations, the effect of this diquark
condensation (formation of quark Cooper pairs) on the EoS is
similar to the occurrence of bound states and corresponds to a
negative pressure contribution (Fig. \ref{fig:bagnew}).

For phenomenological applications of the quark matter EoS (\ref{press})
we provide a polynomial fit of the bag pressure
\begin{eqnarray}
\label{fitbag}
B^{(s)}(\mu_B)= \left\{
\begin{array}{r@{~~}l}\sum_{k=0}^{10} a^{(s)}_k (\mu_B - \mu^{(s)}_c)^k,
& \mu_B > \mu^{(s)}_c  \\
P_{id}(\mu_B), & \mu_B < \mu^{(s)}_c
\end{array} \right..
\end{eqnarray}

The coefficients $a^{(s)}_k$ as well as the critical
chemical potential $\mu^{(s)}_c$ of the chiral phase transition depend on
the choice of the formfactor (see Tabs. \ref{coeff1}, \ref{coeff2}).

The dependence of the diquark gap on the chemical potential can be
represented in a similar way by the polynomial fit
\begin{eqnarray}
\label{fitdelta}
\Delta(\mu_B)= \left\{
\begin{array}{r@{~~}l}\sum_{k=0}^{6} b_k (\mu_B - \mu^{(S)}_c)^k
&,~~ \mu_B > \mu^{(S)}_c \\
0 &,~~ \mu_B < \mu^{(S)}_c
\end{array} \right.~,
\end{eqnarray}
where the coefficients  $b_k$ for the different formfactors
are given in Tab. \ref{coeffdiq}.

The volume fraction also can be approximated by  polynomials in the
following form
\begin{eqnarray}
\label{fitchi}
\chi(\mu_B)= \left\{
\begin{array}
{r@{~~}l}\sum_{k=0}^{6} c_k (\mu_B - \mu^{(\chi)}_c)^k
  &,~~ \mu_B > \mu^{(\chi)}_c  \\
\sum_{k=0}^{1} c_k (\mu_B - \mu^{(S)}_c)^k
  &,~~ \mu^{(\chi)}_c >\mu_B > \mu^{(S)}_c  \\
0 &,~~ \mu_B < \mu^{(S)}_c
\end{array} \right.~~~.
\end{eqnarray}

The coefficients $c_k$ and the chemical potentials $\mu^{(\chi)}_c$ are given 
in the Table \ref{coeffvolfrac} for different formfactors.

\section{Hadronic equation of state and phase transition}

\begin{figure}[hth]
  \begin{center}
    \vspace*{0.7cm}
    \includegraphics[width=0.65\linewidth,angle=-90]{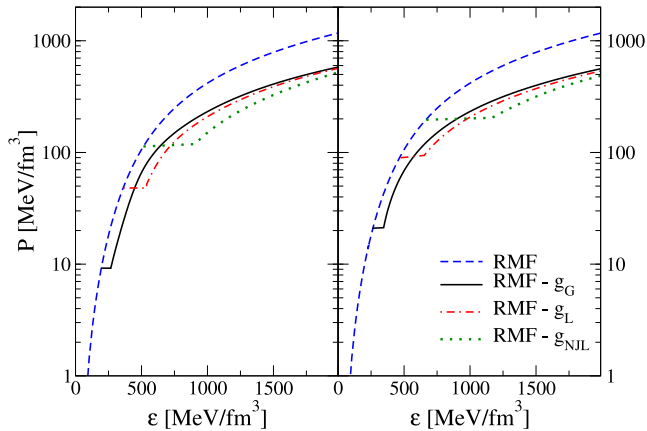}
    \caption{EoS for strongly interacting matter at zero temperature under
compact stars constraints for the coupling parameter $\eta=1$ (left panel) and $\eta=0$ (right panel).
Dashed line: relativistic mean-field model for hadronic matter; solid,
dash-doted and doted lines correspond to quark matter with Gaussian, 
Lorentzian and NJL formfactor functions, respectively.}
    \label{fig:Eos_2in1}
  \end{center}
\end{figure}

At low densities, quarks will be confined in hadrons and an appropriate
EoS for dense hadron matter has to be chosen. For our discussion of
quark-hadron hybrid star configurations in the next Section, we use
the relativistic mean field (RMF) model of asymmmetric nuclear matter
including a non-linear scalar field potential and the $\rho$ meson
(nonlinear Walecka model), see \cite{Glendenning:wn}.
The quark-hadron phase transition is obtained using
the Maxwell construction, see Refs. 
\cite{Voskresensky:2001jq,Voskresensky:2002hu} for a discussion.
The resulting EoS is shown in Fig. \ref{fig:Eos_2in1} 
for the case $\eta=1$ (left panel) when the quark matter phase is superconducting and for $\eta=0$ (right panel) when it is normal.

 \begin{figure}[htb]
  \begin{center}
    \includegraphics[width=0.65\linewidth,angle=-90]{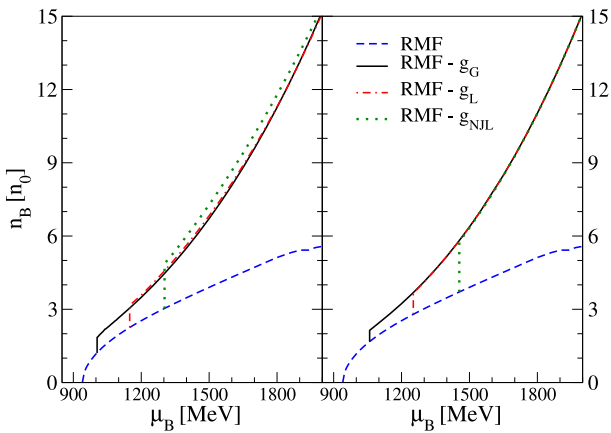}
    \vspace*{0.7cm}
    \caption{Baryon number density in units of the nuclear saturation 
density as a function of baryon chemical potential
 $\mu_B$. Left panel: $\eta=1$. Right panel:  $\eta=0$.
Line styles correspond to the previous figure.}
    \label{fig:Nb_2in1}
  \end{center}
\end{figure}

When comparing the three quark model formfactors under consideration, the 
hardest 
quark matter EoS is obtained for the Gaussian, 
and therefore the critical pressure and
corresponding critical energy densities
of the deconfinement
transition are the smallest, see Table \ref{coeff1}. 
The same statement holds for the case $\eta=0$, when the quark matter 
phases are normal, see right panel of Fig. \ref{fig:Eos_2in1}.

Acording to the Maxwell construction of the deconfinement phase transition, 
there is a jump in the energy density, as is showwn in Fig. \ref{fig:Eos_2in1}.

The corresponding jumps in the baryon densities at the critical chemical 
potentials $\mu_B^{(H)}$ are given in Table \ref{crtdens}, see also Fig. 
\ref{fig:Nb_2in1} for the behavoir of $n_B(\mu_B)$ for all three formfactors and 
both cases of the diquark coupling, $\eta=1$ (left panel) and $\eta=0$ 
(right panel).

The EoS of hybrid stellar matter for temperature $T=0$  is
relevant also for calculations of compact star cooling, since
the star structure is insensitive to the
temperature evolution for $T<1$ MeV.

\section{Configurations of hybrid stars}

\begin{figure}[htb]
  \begin{center}
    \vspace*{0.7cm}
    \includegraphics[width=0.65\linewidth,angle=-90]{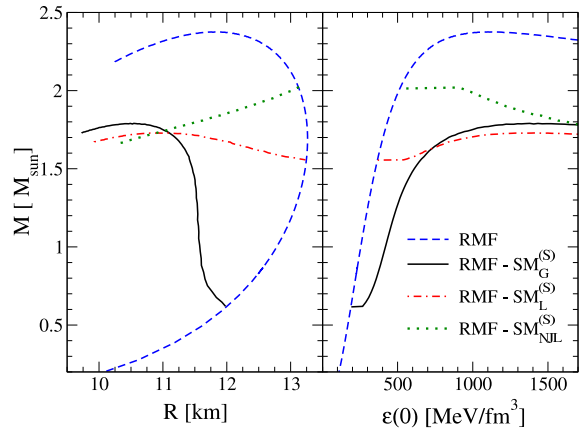}
    \caption{Mass - radius and mass - central energy density relation
for compact stars configurations acording to the EoS shown in the left panel of Fig. \ref{fig:Eos_2in1}.
Hybrid stars with Gaussian or Lorentzian quark matter models give stable 
branches.}
    \label{fig:ConfFF}
  \end{center}
\end{figure}

\begin{figure}[htb]
  \begin{center}
    \vspace*{0.7cm}
    \includegraphics[width=0.65\linewidth,angle=-90]{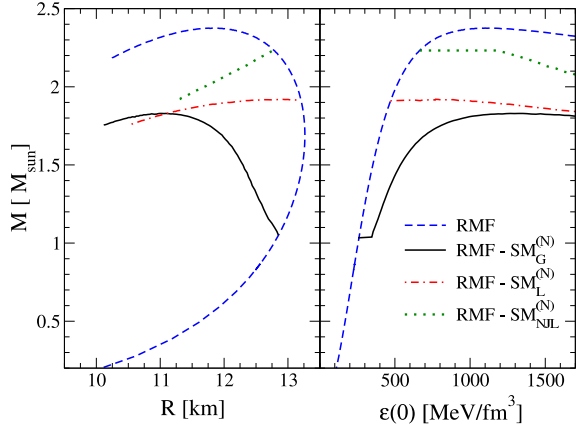}
    \caption{Same as Fig. \ref{fig:ConfFF} for the EoS of the right panel of 
Fig. \ref{fig:Eos_2in1}. }
    \label{fig:Conf_n_FF}
  \end{center}
\end{figure}

\begin{figure}[htb]
  \begin{center}
    \vspace*{0.7cm}
    \includegraphics[width=0.65\linewidth,angle=-90]{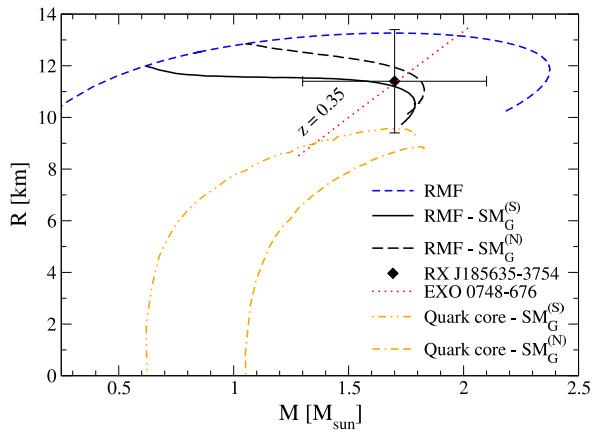}
    \caption{Radius-mass relation for Gaussian formfactor
including the mass dependence
of the quark core radius in both cases, one with and
the other without 2SC phase.}
    \label{fig:Conf_add_1}
  \end{center}
\end{figure}

\begin{figure}[htb]
  \begin{center}
    \vspace*{0.7cm}
    \includegraphics[width=0.65\linewidth,angle=-90]{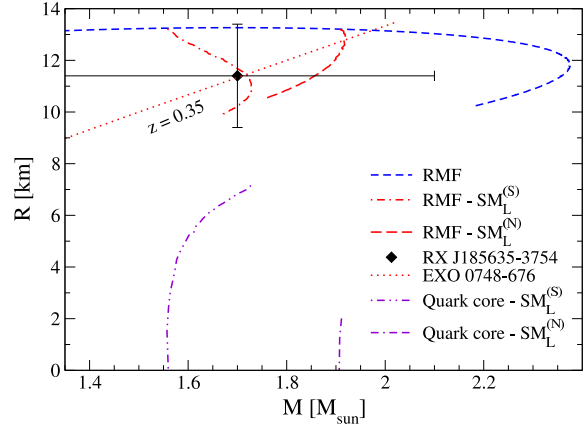}
    \caption{Same as Fig. \ref{fig:Conf_add_1} for Lorentzian formfactor.}
    \label{fig:Conf_add_2}
  \end{center}
\end{figure}

In this Section we consider the problem of stability of cold
($T=0$) hybrid stars with color superconducting quark matter core.
The star configurations are defined from the well known
Tolman-Oppenheimer-Volkoff equations
\cite{Oppenheimer:1939ne}, written  for the hydrodynamical equilibrium of
a spherically distributed matter fluid in General Relativity,
see also \cite{Glendenning:wn},
\begin{equation}
\frac{dP(r)}{dr}=
-\frac{[\varepsilon(r)+P(r)][m(r)+4\pi r^{3}P(r)]}{r[r-2m(r)]},
\end{equation}
where the mass enclosed in a sphere with distance $r$ from the center of
configurations is defined by
\begin{equation}
m(r)=4\pi \int_{0}^{r}\varepsilon(r')r'^{2}dr'.
\end{equation}
These equations are solved for a set of central energy
densities, see Figs. \ref{fig:ConfFF} -  \ref{fig:Conf_add_2}.
An approximate criterion for the stability of star configurations
is that masses should be rising functions
of the central energy density $\varepsilon(0)$.


Our calculations show that for the Gaussian and Lorentzian
formfactors one can have stable configurations with a quark core,
either with  (Fig. \ref{fig:ConfFF}) or  without (Fig. \ref{fig:Conf_n_FF})
color superconductivity
whereas for our parameterization of the NJL model (cutoff formfactor) the
configurations with quark cores are not stable.
For the Gaussian formfacor case the occurence of color
superconductivity in quark matter shifts the critical mass of the hybrid star 
from $1.04$ M$_\odot$ to  $0.62$ M$_\odot$ and
the maximal value of the hybrid star mass from $1.83$ M$_\odot$ to
$1.79$ M$_\odot$.
For the Lorentzian formfactor the branch of stable hybrid stars with
2SC supercondcting quark cores lies in the mass range between 
$1.55$ M$_\odot$ and the maximum mass $1.72$ M$_\odot$.


In Figs.\ref{fig:Conf_add_1} and \ref{fig:Conf_add_2} we demonstrate
that these
models fulfill the observational constraints from the isolated neutron star
RX J$185635-3754$ \cite{Prakash:2002xx,Zane:2003bp} and from
the observation of the surface redshift for EXO 0748-676 \cite{Cottam:2002cu}.

The Lorentzian model with normal quark matter has marginally 
stable quark cores with radii less than $2$ km, in the mass range 
$1.91$ -$1.92 ~M_\odot$, see Fig. \ref{fig:Conf_add_2}.

\section{Conclusion}

We have investigated the influence of the diquark condensation on
EoS of quark matter and obtained the critical densities of phase
transition to hadronic matter for different formfactors of quark
interaction.

We find that the charge neutrality condition requires that the quark matter
phase consists of a mixture of 2SC condensate and normal phase. The volume fraction of the condensate phase amounts to 65\% - 85\% depending on the
formfactor function of the interaction. In the present work we did not
consider muons in the quark matter phase. Their  occurence would
increase the volume fraction of the superconducting phase by
about a  5\%, helping to
stabilize the 2SC phase.

We have shown that for our set of formfactors the NJL model
gives no stable quark core hybrid stars. 
The occurence of the superconducting 2SC phase in quark matter supports 
the stability of the quark matter phase.

Comparison of the quark core neutron star mass-radius relation with
the mass and radius of the recently observed 'small' compact object
RX J$185635-3754$ and with the constraints from the observation of the 
surface redshift for the object EXO 0748-676 shows that our model perfectly 
obeys those constraints.

These studies can be viewed as a preparatory step before more fundamental
nonperturbative interactions can be provided, e.g. by QCD Schwinger-Dyson
Equation studies \cite{Bender:1996bm,Roberts:2000aa,Maas:2002if}.

\section*{Acknowlegements}
We thank our colleagues for discussions and interest in our work, in 
particular during the NATO workshop in Yerevan, Armenia.
Special thanks go to M. Buballa and D. Rischke for important remarks
on a previous version of this work.

The research of D.N. Aguilera has been supported by DFG Graduiertenkolleg 567
``Stark korrelierte Vielteilchensysteme'', by  CONICET PIP 03072 (Argentina), 
by DAAD grant No. A/01/17862 and by the Harms Stiftung of the University 
of Rostock.
H.G. acknowledges support by DFG under grant No. 436 ARM 17/5/01 and by the
Virtual Institute of the Helmholtz Association ``Dense Hadronic Matter and QCD 
Phase Transitions'' under grant No. VH-V1-041.


\begin{table}[H]
\begin{tabular}{|c||c|c|c||c|c|c|}\hline
 Form &$\Lambda$&$G_1~\Lambda^2$&$m$&$T_c(\mu=0)$&
$\mu_{c}^{(S)}(T=0)$&$\mu_{c}^{(N)}(T=0)$\\
Factor&[GeV]&&[MeV]&[MeV]&[MeV]&[MeV]\\ \hline
Gauss.   &$1.025 $&$3.7805$&$2.41$&$174$&$965$&$991$\\
Lor. &$0.8937$&$2.436 $&$2.34$&$188$&$999$&$1045$\\
NJL        &$0.9   $&$1.944 $&$5.1 $&$212$&$1030$&$1100$\\ \hline
  \end{tabular}
  \caption{Parameter sets ($\Lambda$, $G_1~\Lambda^2$, $m$) of the nonlocal
chiral quark model for different formfactors discussed in the text.
The last three columns show the critical temperatures at vanishing chemical
potential and the critical chemical potentials with and without diquark
condensate at vanishing temperature, respectively.}
  \label{par}
\end{table}


\begin{table}[H]
  \begin{tabular}{|r|r|r|r|}\hline
$k$&\multicolumn{3}{|c|}{$a_k^{(N)}$ [GeV$^{1-k}$fm$^{-3}$]}   \\
\hline
 &  Gaussian &  Lorentzian  &  NJL  \\
\hline
$0$ & $ 7.2942\cdot 10^{-2}$&$ 9.0218\cdot 10^{-2}$&$ 1.1071\cdot 10^{-1}$\\
$1$ & $ 2.5122\cdot 10^{-2}$&$ 7.6973\cdot 10^{-2}$&$ 3.0219\cdot 10^{-1}$\\
$2$ & $-9.1152\cdot 10^{-2}$&$-6.8728\cdot 10^{-1}$&$ 1.2820\cdot 10^{+0}$\\
$3$ & $ 1.6402\cdot 10^{-1}$&$ 3.7260\cdot 10^{+0}$&$-4.0634\cdot 10^{+1}$\\
$4$ & $-5.9621\cdot 10^{-3}$&$-1.1862\cdot 10^{+1}$&$ 3.0828\cdot 10^{+2}$\\
$5$ & $-5.1899\cdot 10^{-1}$&$ 2.2342\cdot 10^{+1}$&$ 1.2301\cdot 10^{+3}$\\
$6$ & $ 9.0892\cdot 10^{-1}$&$-2.4464\cdot 10^{+1}$&$ 2.9441\cdot 10^{+3}$\\
$7$ & $-6.5617\cdot 10^{-1}$&$ 1.4374\cdot 10^{+1}$&$-4.3677\cdot 10^{+3}$\\
$8$ & $ 1.7810\cdot 10^{-1}$&$-3.5006\cdot 10^{+0}$&$ 3.9376\cdot 10^{+3}$\\
$9$ & $                   0$&$                   0$&$-1.9774\cdot 10^{+3}$\\
$10$& $                   0$&$                   0$&$ 4.2442\cdot 10^{+2}$\\
\hline
\end{tabular}
  \caption{Coefficients for Bag function fit formula for the normal phase case,
for different formfactors  Eq. (\ref{fitbag}).}
  \label{coeff1}
\end{table}

\begin{table}
  \begin{tabular}{|r|r|r|r|}\hline
$k$&\multicolumn{3}{|c|}{$a_k^{(S)}$ [GeV$^{1-k}$fm$^{-3}$]}   \\
\hline
 &  Gaussian &  Lorentzian  &  NJL  \\
\hline
$0$ & $ 6.5168\cdot 10^{-2}$&$ 7.5350\cdot 10^{-2}$&$ 8.4897\cdot 10^{-2}$\\
$1$ & $ 1.4638\cdot 10^{-1}$&$ 2.8766\cdot 10^{-1}$&$ 2.8604\cdot 10^{-1}$\\
$2$ & $-1.8020\cdot 10^{+0}$&$ 3.2215\cdot 10^{+0}$&$ 1.0708\cdot 10^{+0}$\\
$3$ & $ 9.8125\cdot 10^{+0}$&$ 1.6528\cdot 10^{+1}$&$-2.8157\cdot 10^{+1}$\\
$4$ & $-3.0515\cdot 10^{+1}$&$-4.9352\cdot 10^{+1}$&$ 1.6904\cdot 10^{+2}$\\
$5$ & $ 5.4888\cdot 10^{+1}$&$ 8.7023\cdot 10^{+1}$&$-5.4828\cdot 10^{+2}$\\
$6$ & $-5.6610\cdot 10^{+1}$&$-8.9237\cdot 10^{+1}$&$ 1.0896\cdot 10^{+3}$\\
$7$ & $ 3.1096\cdot 10^{+1}$&$ 4.9210\cdot 10^{+1}$&$-1.3643\cdot 10^{+3}$\\
$8$ & $-7.0479\cdot 10^{+0}$&$-1.1275\cdot 10^{+1}$&$ 1.0518\cdot 10^{+3}$\\
$9$ & $                   0$&$                   0$&$-4.5653\cdot 10^{+2}$\\
$10$& $                   0$&$                   0$&$ 8.5443\cdot 10^{+1}$\\

\hline
\end{tabular}
  \caption{Coefficients for Bag function fit formula for the superconducting mixed phase,
for different formfactors, Eq. (\ref{fitbag}).}
  \label{coeff2}
\end{table}
\begin{table}
  \begin{tabular}{|r|r|r|r|}
\hline
$k$&\multicolumn{3}{|c|}{$b_k$ [GeV$^{1-k}$]}   \\
\hline
 &  Gaussian &  Lorentzian  &  NJL  \\
\hline
$0$ &$ 9.33\cdot 10^{-2}$&$ 1.04\cdot 10^{-1}$&$ 3.30\cdot 10^{-2}$\\
$1$ &$ 2.13\cdot 10^{-1}$&$ 1.63\cdot 10^{-1}$&$ 1.05\cdot 10^{0~}$\\
$2$ &$-4.27\cdot 10^{-2}$&$ 9.19\cdot 10^{-2}$&$-3.42\cdot 10^{0~}$\\
$3$ &$ 1.14\cdot 10^{-2}$&$-1.92\cdot 10^{-1}$&$ 6.13\cdot 10^{0~}$\\
$4$ &$-5.27\cdot 10^{-3}$&$ 8.88\cdot 10^{-2}$&$-5.20\cdot 10^{0~}$\\
$5$ &$0                 $&$0                 $&$ 1.55\cdot 10^{0~}$\\
$6$ &$0                 $&$0                 $&$ 1.03\cdot 10^{-1}$\\
\hline
\end{tabular}
\caption{Coefficients for the diquark condensate fit formula, for different formfactors, Eq.
(\ref{fitdelta}).}
\label{coeffdiq}
\end{table}

\begin{table}
  \begin{tabular}{|r|r|r|r|}\hline
$k$ &    \multicolumn{3}{|c|}{$c_k$  [GeV$^{-k}$]}   \\
\hline
    &  Gaussian              &  Lorentzian           &  NJL  \\
\hline
$\mu_{c}^{(\chi)}$[MeV]& $995$          &$1054$                 &$1095$  \\
\hline
&    \multicolumn{3}{|c|}{ for  
$ \mu^{(S)}_c < \mu_B < \mu^{(\chi)}_c$}   \\
\hline
$0$ & $ 1.29\cdot 10^{-1}$&$ 8.56\cdot 10^{-2}$&$ 7.00\cdot 10^{-3}$\\
$1$ & $ 1.66\cdot 10^{+1}$&$ 1.15\cdot 10^{+1}$&$ 8.51\cdot 10^{0~}$\\
\hline
&    \multicolumn{3}{|c|}{ for $ \mu^{(\chi)}_c <\mu_B $}   \\
\hline

$0$ & $ 6.28\cdot 10^{-1}$&$ 7.17\cdot 10^{-1}$&$ 5.60\cdot 10^{-1}$\\
$1$ & $ 3.36\cdot 10^{-1}$&$ 2.83\cdot 10^{-1}$&$ 2.47\cdot 10^{0~}$\\
$2$ & $-3.94\cdot 10^{-1}$&$-3.43\cdot 10^{-1}$&$-7.67\cdot 10^{0~}$\\
$3$ & $ 6.57\cdot 10^{-2}$&$-1.19\cdot 10^{-2}$&$ 1.06\cdot 10^{+1}$\\
$4$ & $-8.78\cdot 10^{-3}$&$ 2.50\cdot 10^{-2}$&$-5.34\cdot 10^{0~}$\\
$5$ & $                 0$&$                 0$&$-6.25\cdot 10^{-1}$\\
$6$ & $                 0$&$                 0$&$ 7.45\cdot 10^{-1}$\\
\hline
\end{tabular}
  \caption{Coefficients for the volume fraction Eq. (\ref{fitchi}) and their valid ranges,
for different formfactors.}
  \label{coeffvolfrac}
\end{table}

\begin{table}
  \begin{tabular}{|c|c|c|c|c|c|c|}\hline
             &\multicolumn{2}{|c|}{$n_B^{(Q)}[n_0]$}
             &\multicolumn{2}{|c|}{$n_B^{(H)}[n_0]$}
             &\multicolumn{2}{|c|}{$\mu_B^{(H)}[{\rm MeV}]$}\\ \hline
    &$\eta = 1$&$\eta=0$&$\eta = 1$&$\eta=0$&$\eta = 1$&$\eta=0$\\
\hline
    Gaussian   &$1.84$ & $2.14$ & $1.20$ & $1.68$ &$1005$& $1059$  \\
    Lorentzian &$3.16$ & $3.66$ & $2.25$ & $2.79$ &$1149$& $1252$   \\
    NJL        &$4.79$ & $5.76$ & $3.02$ & $3.71$ &$1303$& $1455$    \\ 
\hline
  \end{tabular}
  \caption{Limiting densities of the coexistence region between
 quark (Q) and hadron (H) matter phases for different
formfactors  ($n_0 = 0.16~{\rm fm}^{-3}$ 
is the nuclear saturation density) in the first two columns. The third 
column shows the critical baryon chemical potential at the phase transition
 $\mu_B^{(H)}=\mu_B^{(Q)}$, see Fig. \ref{fig:Nb_2in1}. The subcolumns
indicate 
the cases of superconducting ($\eta=1$) and normal ($\eta=0$) quark matter.}
  \label{crtdens}
\end{table}

\end{document}